# Continuous Wave Second Harmonic Generation from an Etchless Lithium Niobate Resonant Metasurface

*Zetian Chen\*, Noa Mazurski, and Uriel Levy\**

*The Faculty of Science, The Center for Nanoscience and Nanotechnology, Institute of Applied Physics, The Hebrew University of Jerusalem, Jerusalem 91904, Israel*

*E-mail: zetian.chen@mail.huji.ac.il; ulevy@mail.huji.ac.il*

**Abstract:** Nonlinear metasurfaces provide a route to compact frequency conversion by replacing phase matching and long interaction lengths with resonantly enhanced light–matter interaction in subwavelength structures. Extending this capability to continuous-wave (CW) operation is particularly important for applications requiring narrow linewidth, stable frequency, and stationary optical fields, but remains extremely challenging. Here, we demonstrate CW second-harmonic generation on a transmission-mode, etchless thin-film lithium niobate platform enabled by a patterned silicon-rich nitride metasurface. This hybrid design combines guided-mode-resonance coupling, low optical loss, and CMOS-compatible processing while keeping most of the optical mode confined in the unpatterned lithium niobate, yielding a measured quality factor of ~2300. Clearly resolved SHG is achieved under sub-kW/cm² CW pumping, with a normalized conversion efficiency of 0.156 % cm²/GW in the low-power regime. Interestingly, our work reveals that CW resonant SHG in metasurfaces can exhibit pronounced transient dynamics, including power-dependent resonance evolution, overshoot, and nonideal scaling. These findings establish etchless LN-SRN metasurfaces as a promising platform for compact CW nonlinear photonics, and show that resonance dynamics are central to the operation and evaluation of CW-driven nonlinear metasurfaces.

## 1. Introduction

Second Harmonic Generation (SHG) is a nonlinear optical process[1] in which two photons of the same frequency interact with a material to produce a new photon at twice the frequency. It is widely used in laser frequency doubling[2], surface and molecular spectroscopy[3], and nonlinear imaging[4–6]. In integrated photonics, efficient SHG is most commonly realized in quasi-phase-matched waveguide platforms[7–9]. Although highly effective, this approach generally requires careful dispersion and polarization engineering over sufficient propagation length, which constrains device geometry, excitation conditions, and fabrication tolerance. Resonant dielectric metasurfaces provide a fundamentally different route to SHG. By exploiting resonance-induced field enhancement, they can drive nonlinear conversion without conventional phase matching and enable SHG from subwavelength films under free-space excitation[10–12]. This makes metasurfaces attractive for compact nonlinear photonics[13]. Yet the material choices remain limited. Silicon, despite its maturity in photonic integration, does not support bulk $\chi^{(2)}$. Other efficient $\chi^{(2)}$ materials, such as GaP[14] and lithium niobate (LN)[15–18], are less naturally aligned with standard CMOS-style nanofabrication when direct patterning of the nonlinear medium is required. This has motivated etchless LN metasurface concepts, in which the LN layer remains intact, and the optical resonance is introduced through a patterned capping layer with a refractive index close to that of LN[19–25]. In addition, nonlinear metasurface SHG has been demonstrated predominantly under pulsed excitation[12,26]. By contrast, many applications require CW-pump nonlinear conversion[14,27], especially when narrow linewidth, stable frequency, coherent phase, and stationary fields are needed, such as laser frequency stabilization using narrow atomic transitions at the up-converted frequency[9,28], as well as narrowband spectroscopy and precision metrology[29–31].

Here, we demonstrate CW-pumped SHG on an etchless thin-film LN (TFLN) platform using a patterned silicon-rich nitride (SRN) metasurface (Figure 1a). The thin SRN layer with low Si concentration enables guided-mode-resonance (GMR) coupling while keeping most of the optical mode inside the unpatterned LN layer (Figure 1b), yielding a measured quality (Q) factor of ~2300. Similar to $TiO_2$[21,25], SRN can provide low optical loss in the visible and near-infrared; at the same time, it can be deposited and patterned by CMOS-compatible processes[32]. With this hybrid LN-SRN design, we observe clearly measurable SHG under sub-kW/cm² CW pumping, whereas an unpatterned region of the same sample remains below the detection limit even at much higher pump power. A normalized conversion efficiency of 0.156 % cm²/GW is obtained in the low-power regime, which is highly competitive among the etchless LN SHG metasurfaces. Furthermore, this platform also allows us to resolve the transient behavior of CW resonant SHG, including power-dependent resonance evolution, overshoot, and nonideal scaling (Figure 1a). These observations show that CW nonlinear conversion in resonant metasurfaces is governed not only by static field enhancement, but also by the dynamic evolution of the resonant state under illumination.

## 2. Results
### 2.1. Sample

The device structure, operating concept, and key observed behavior are illustrated in Figure 1. The device is built on a commercially available lithium niobate on insulator (LNOI) platform, consisting of an approximately 310-nm-thick x-cut TFLN layer on $SiO_2$. Thin silicon nitride strips (with period of 1.8 um in x-direction, thickness of 80 nm and width w=1.2 um) are deposited and patterned on top of the LN using standard CMOS-compatible PECVD and RIE processes. By slightly increasing the silicon precursor concentration during deposition, the resultant Si-rich nitride (SRN) appears to have refractive index close to that of lithium niobate in the telecom C-band ($n_{LN}(e)$~2.14, and $n_{SRN}$~2.18) while still maintaining minimal loss in both fundamental and SH wavelengths (Figure S1). This small index contrast allows the optical mode to be smoothly distributed across both materials. In principle, such a hybrid structure can support a fundamental transverse-electric (TE) mode (x-polarized) propagating in the $xy$ plane, as confirmed by the eigen mode simulation in Figure 1b. Nearly 70% of the modal power is confined in the LN layer, owing to the much smaller thickness of the SRN layer. With the optical polarization aligned along the extraordinary axis of LN, this configuration can access the largest second-order nonlinear coefficient of LN[33] and thereby favor efficient SHG, as described by the equation below:

$$P_Z(2\omega) \approx 2\epsilon_0 d_{33} E_Z^2(\omega) \qquad (1)$$

Where $P$ is the second order nonlinear polarization, $d_{33}$ is the largest nonlinear susceptibility tensor element of LN (~-25 pm/V in near infrared), and $E$ is the optical electric field at fundamental wavelength (FW). $Z$ refers to the extraordinary optical axis direction (x axis in Figure 1).

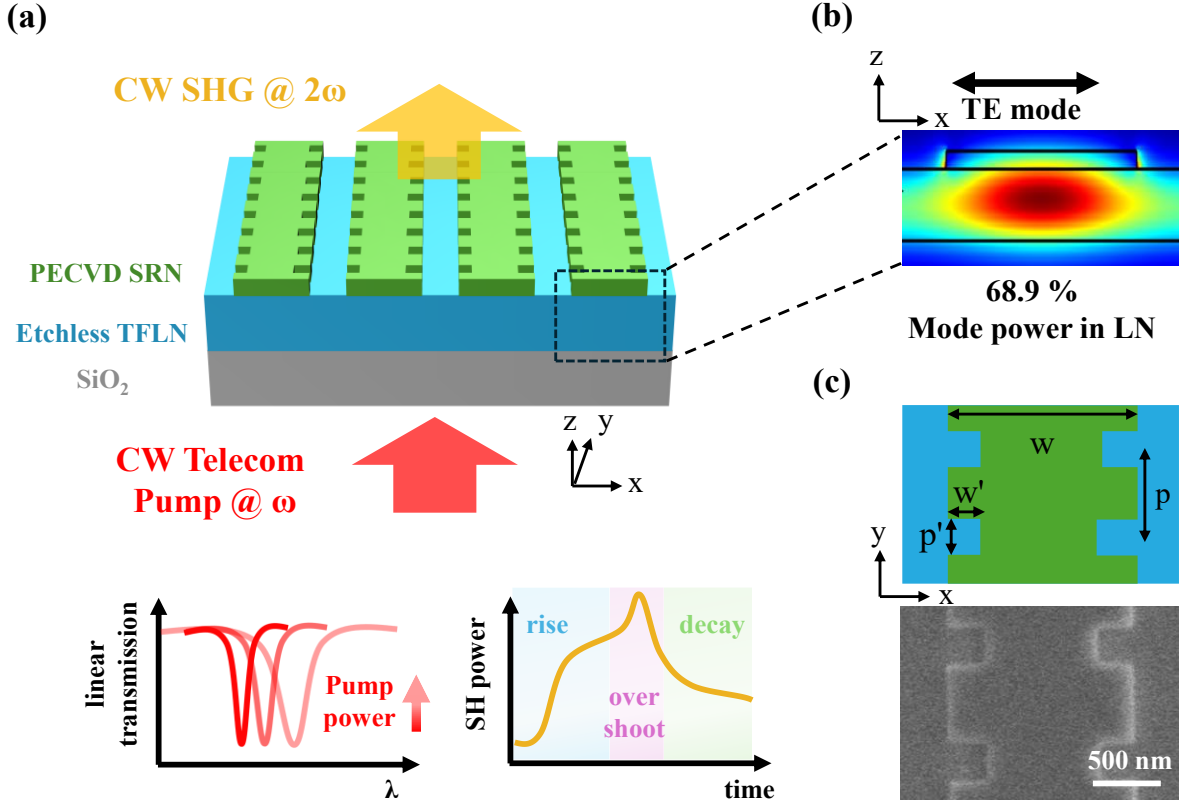

Figure 1. Overview of the device structure, operating concept, and key observed behavior. (a) Schematic of the etchless thin-film lithium niobate (TFLN) metasurface, consisting of a 310-nm-thick TFLN layer on insulator and a PECVD-deposited silicon-rich nitride (SRN) metasurface on top. A CW telecom pump is incident from the substrate side, and the generated CW second harmonic is collected in transmission. As discussed in the main text, the device exhibits pump-power-dependent resonance evolution, including red shift and apparent broadening, together with transient SH dynamics featuring rise, overshoot, and decay. (b) Simulated eigenmode profile of a unit cell (1.8 μm along x), showing a fundamental TE mode strongly confined in the unpatterned LN layer. (c) Top view of the notched SRN metasurface, where the periodic perturbation enables excitation of the guided-mode resonance associated with the mode in (b). Bottom: SEM image of the fabricated device.

To couple light from free space into this in-plane guided mode, periodic notches are introduced along the SRN bar[34,35], as seen from the device top view in Figure 1c, with the corresponding fabricated structure presented in the SEM image below. These periodic perturbations enable excitation of guided-mode resonances (GMRs), whose resonance wavelength is approximately determined by:

$$p = \lambda_{eff,i} = \lambda_0/n_{eff,i} \qquad (2)$$

where p is the notch period, $\lambda_{eff,i}$ is the effective guided mode wavelength of the $i$th-order GMR, $n_{eff,i}$ is the corresponding effective refractive index, and $\lambda_0$ is the free-space wavelength. For the mode shown in Figure 1b ($n_{eff,1}$=1.8), a period of $p \sim 860$ nm is therefore selected to place the GMR wavelength near 1550 nm. The notch size, which defines the perturbation strength, governs the coupling efficiency and correspondingly the quality factor (Q factor) of the resonance. Under normal incidence along the SRN bar with polarization in x-direction, the simulated transmission spectra in Figure 2a show that near-critical coupling with $Q \sim 2700$ can be achieved for a notch size of $w' = 240$ nm and $p' = 260$ nm (hereafter referred to as the high-Q device). Because the SRN layer is thin, this geometry still provides a practically accessible perturbation size for fabrication. A larger notch ($w' = 360$ nm and $p' = 430$ nm) produces a lower Q factor of $\sim 800$ (hereafter referred to as the low-Q device). In addition, varying the notch period allows the resonance to be positioned at different wavelengths around 1550 nm. We note that only the zeroth-order transmitted light is monitored here. The weak diffraction into higher orders (in x-direction, cross the SRN bars), indicated by the high off-resonant zeroth-order transmission, is a consequence of the low-index-contrast grating formed by the thin SRN layer[36].

The inset of Figure 2a shows the calculated optical electric field of the high-Q device at resonance (cross section in the xz-plane at the middle of the notch). A strong local electric-field enhancement is observed within the LN region (white dashed box), reaching up to 44-fold in field amplitude at resonance, which corresponds to an intensity enhancement of approximately $2 \times 10^3$. Such strong resonant field buildup, together with the narrow linewidth, is favorable for SHG under continuous-wave (CW) excitation, where efficient spectral overlap between the pump laser and the resonance can be maintained. This contrasts with many pulsed-excitation conditions, in which the pump bandwidth can exceed the resonance linewidth and reduce the effective resonant enhancement.

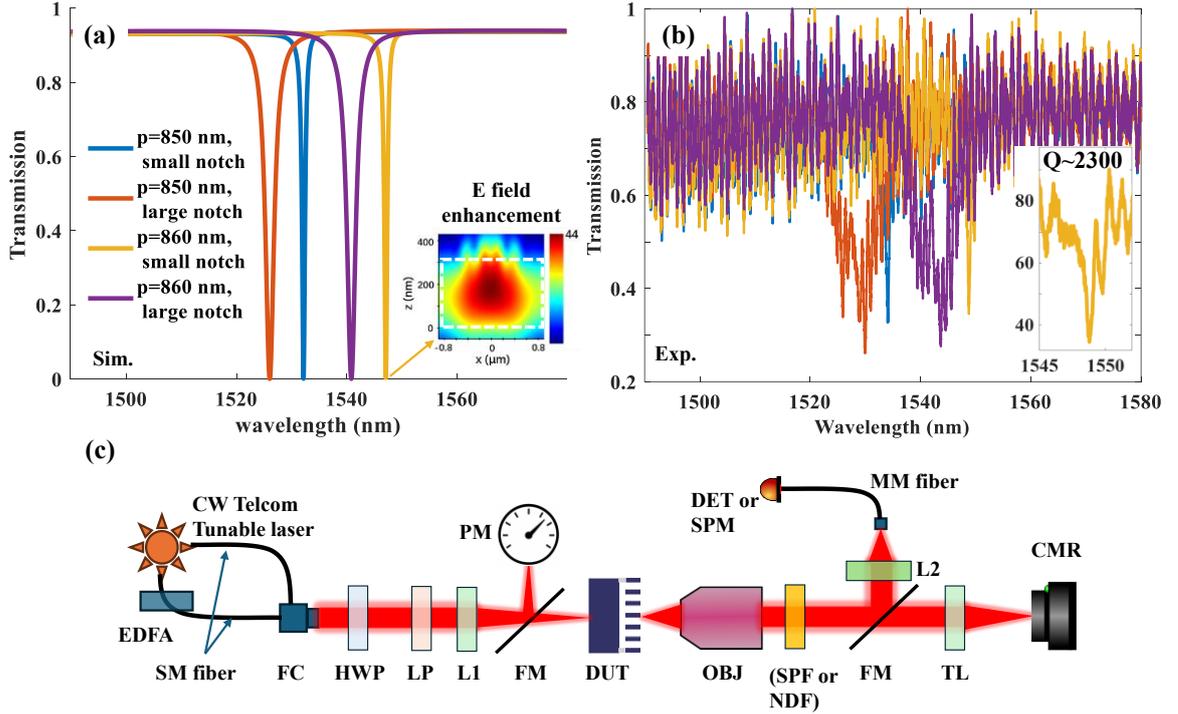

*Figure 2. Linear transmission spectrum. (a) Simulated spectrum of devices with different notch sizes and periods. Inset: calculated field enhancement for the small notch high-Q device at resonance, showing 44-fold enhancement confined in the LN. (b) Measured transmission spectrum. Inset shows the zoom-in of the high-Q device resonance, estimated with Q~2300. (c) Optical setup. EDFA: erbium-doped fiber amplifier, SM: single mode, FC: fiber collimator, HWP: half-wave plate, LP: linear polarizer, L: lens, FM: flip mirror, PM: power meter, DUT: device under test, OBJ: objective, SPF: short-pass filter (for SH acquisition), NDF: neutral-density filter (for transmission spectrum measurement under elevated power), MM: multi-mode, DET: detector, SPM: spectrometer, TL: tube lens, CMR: camera.*

## 2.2 Optical characterizations

We next turn to the optical characterization of the fabricated devices. The experimental setup is schematically illustrated in Figure 2c. For the transmission measurements in the FW spectral range, a continuous-wave (CW) telecom tunable laser is used as the light source. The laser output is delivered to a fiber collimator (FC) through a single-mode (SM) fiber. After passing through a half-wave plate (HWP) and a linear polarizer (LP), the beam is weakly focused onto the device using a long-focal-length lens (L1). The zeroth-order transmitted beam is collected by a long-working-distance 4× objective (OBJ, NA = 0.13). The collected light is then coupled into a multimode (MM) fiber, which is connected either to the built-in detector of the tunable laser for transmission-spectrum measurements or to a camera for device alignment.

The measured transmission spectra are shown in Figure 2b. Each curve is normalized to its own peak transmission. Although Fabry–Pérot oscillations arising from multiple reflections in the thick substrate are superimposed on the spectra, distinct resonances are clearly resolved. The measured resonance wavelengths agree reasonably well with the simulated values. No observable difference in the transmission spectrum is found for illumination from either side of the sample, in either simulation or experiment. The measured Q factors are estimated to be approximately 2300 for the high-Q device and approximately 450 for the low-Q device. Both devices exhibit pronounced resonance contrast, indicating efficient excitation of the resonant mode.

After identifying the FW resonances, we proceed to measure the nonlinear conversion of the devices under increased pump power. For this purpose, an erbium-doped fiber amplifier (EDFA) is inserted between the tunable laser and the FC, with SM fibers used at both the input and output of the amplifier (Figure 2c). The remainder of the setup is kept identical to that used for the linear transmission measurements, thereby maintaining the same alignment and illumination geometry. The HWP is used to control the incident power, which is monitored by a power meter placed immediately before the sample. In the following, we focus on the nonlinear response of the low-Q and high-Q devices with a notch period of $p = 860$ nm, whose resonance wavelengths lie within the optimal operating range of the EDFA.

We first measured the linear transmission spectra under elevated pump powers. Neutral-density (ND) filters are inserted after the OBJ to reduce the optical power returned to the detector. Figure 3a and b show the measured spectra illuminated from the substrate side of the high-Q and low-Q devices, respectively. In both cases, the resonances undergo a red shift as the pump power increases. Because the SM fibers connected to the EDFA are not polarization-maintaining, whereas the input polarization at the sample is fixed by the LP placed immediately before the sample, power fluctuations occur during the wavelength sweep. These fluctuations appear as a reduction in transmission near the edges of the scanned spectral range, compared with the low-power sweep performed without the EDFA (blue curve). For each power level, the incident power was calibrated near the resonance wavelength by rotating the HWP to align the EDFA output polarization with the LP. As shown in Figure 3a, the high-Q device exhibits resonance red shifts of approximately 0.25 nm at 170 mW and 0.86 nm at 460 mW. In contrast, the low-Q device shows smaller red shifts, estimated to be approximately 0.2 nm and 0.7 nm, respectively. The shift is more pronounced in the high-Q device not only in absolute wavelength change, but also relative to the resonance linewidth. Under 170 mW, the resonance dip of the high-Q device is already clearly displaced, and under 460 mW the shift exceeds the full width at half maximum (FWHM) of the resonance. By comparison, for the low-Q device the total red shift remains small relative to its broader linewidth even at 460 mW, as seen from the zoomed-out spectrum in the inset of Figure 3b. In addition, the resonance red shift becomes smaller when the scan speed is increased from 0.5 nm/s to 5 nm/s, as observed for the high-Q device in Figure 3a.

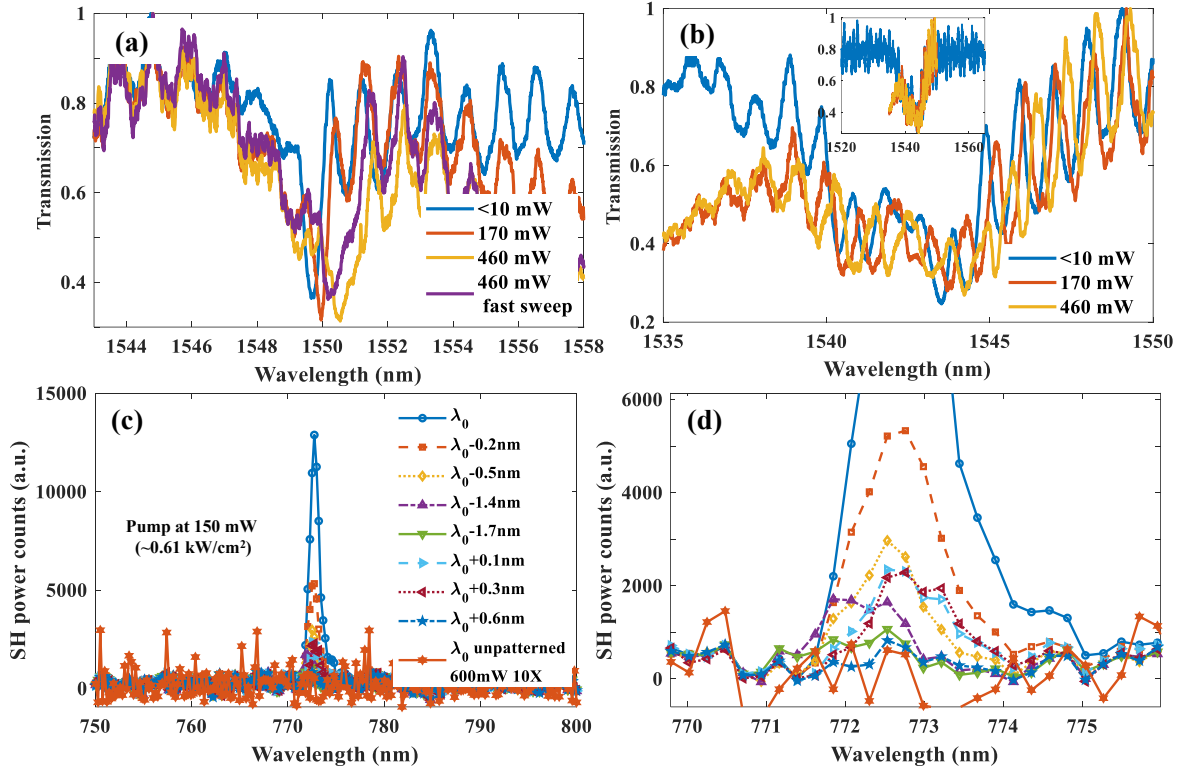

*Figure 3. Linear transmission spectrum shifts under different pump power for (a) high-Q device, and (b) low-Q device (inset is the zoom-out of the resonance, showing shift is marginal compared to its resonance width). (c) Measured SH spectrum of the high-Q device under 150 mW (~0.61 kW/cm²) pump around the resonance. $\lambda_0$ (1550.06 nm) corresponds to the pump wavelength results in the maximum quasi-steady state SH signal under such power. (d) Zoom-in of (c), showing the noise-level signal from an unpatterned area even with higher pump power and longer integration time, as well as the SH wavelength shift following the pump wavelength swing.*

To ensure a consistent comparison of SHG across different pump powers, we adopted a resonance-tracking measurement protocol rather than relying directly on the high-power FW transmission sweep. Efficient SHG in these metasurfaces relies on resonant field enhancement, with the largest enhancement expected at resonance according to the simulations. Because the resonance wavelength shifts with pump power and depends on the scan speed, the exact resonance condition under a given pump power cannot be reliably inferred from a simple FW transmission sweep. At the same time, a meaningful comparison of nonlinear conversion across different pump powers requires the device to be characterized under the same resonant condition, ideally at or near the instantaneous resonance. This consideration is particularly important for CW excitation, where the optical energy is presumably concentrated at a single frequency. Moreover, since the conversion in our device is governed by resonance-enhanced field buildup rather than phase matching, a slight change in pump wavelength is not expected to fundamentally alter the conversion efficiency, provided that the pumping condition remains near resonance. Accordingly, we characterize the SHG response as follows. The ND filters are firstly replaced with short-pass filters (SPFs) to suppress the residual FW pump. For a given pump power, the pump wavelength is initially set slightly to the short-wavelength side of the resonance, outside the resonant dip. The pump wavelength is subsequently increased slowly over a narrow spectral window, where the power fluctuation is negligible, until the resonance is approached in small steps. At the same time, the spectrum at the upconverted wavelength is monitored by connecting the MM fiber to a spectrometer. After each wavelength adjustment, the SH signal is allowed to stabilize before the next step is taken, and its level is compared with that measured at the previous wavelength. The pump wavelength is then iteratively tuned back and forth around the resonance until the SH signal reaches a maximum at quasi-steady state.

Figure 3c shows the measured SH spectrum of the high-Q device under a pump power of 150 mW, corresponding to an incident intensity of only ~0.61 kW/cm$^2$. A clear SH peak is observed from the spectrometer at ~773 nm when the device is pumped at $\lambda_0$=1550.06 nm set on the tunable laser. The slight deviation of the SH peak from exactly half of the FW wavelength, together with its apparent spectral broadening, is mainly attributed to the limited wavelength accuracy and spectral resolution of the spectrometer used in this work, as well as to nonperfect calibration of the laser source. The SH response was further measured at several pump wavelengths in the vicinity of $\lambda_0$, by tuning the pump to both the shorter- and longer-wavelength sides, as shown in the zoomed-in plot in Figure 3d. As the pump wavelength slightly moves away from the optimum resonance condition, the SH signal decreases rapidly.

The SH intensity exhibits an asymmetric sensitivity to detuning on the two sides of the resonance. A red detuning of only ~0.6 nm from $\lambda_0$ is sufficient to reduce the SH signal to the noise level, whereas a blue detuning of approximately ~1.7 nm is required to produce a similar reduction. In principle, the SH wavelength should also be continuously tunable along with the pump wavelength, although in the present measurements this shift is only weakly resolved, again because of the limited spectrometer resolution. Such one-to-one wavelength tracking is more directly accessible under narrowband CW pumping than under broadband pulsed excitation. For comparison, the SH signal was also measured from an unpatterned region of the same sample. No discernible SH spectral feature is observed, even when the pump power is increased to 600 mW and the spectrometer integration time is extended by a factor of 10, which reaches the maximum allowed by the instrument. We note that the sample is based on an LNOI stack, in which the 310-nm TFLN layer is separated from the LN handle substrate by a $SiO_2$ layer. The absence of measurable SHG from the unpatterned region indicates that, under our experimental conditions, neither the thin-film region nor the underlying LN structure produces an observable SH response without resonant enhancement. These results confirm that the detected SHG is dominated by the metasurface-induced field concentration, enabling clearly measurable CW-driven SHG from an unetched 310-nm-thick LN layer at sub-kW/$cm^2$ pump intensity.

We next quantitatively characterize SH conversion as a function of pump power. The pump power is controlled using the EDFA output setting in combination with the HWP. For these measurements, the MM fiber is connected to a silicon femtowatt detector. At each pump power, the SH signal is recorded using the resonance-tracking procedure described above. Figure 4 summarizes the pump-power dependence of the SH power (Figure 4a, c) and the corresponding conversion efficiencies (Figure 4b, d) for the low-Q (Figure 4a, b) and high-Q (Figure 4c, d) devices, respectively. The pump wavelength corresponding to each data point is shown in Figure S2. For the low-Q device, the SH power follows a nearly quadratic dependence on pump power up to ~700 mW, with a fitted slope of 2.06 and a coefficient of determination of $R^2$~0.999 in the log-log plot. This behavior is consistent with the expected power scaling of SHG. Correspondingly, the normalized conversion efficiency remains approximately constant at ~0.01 % $cm^2$/GW over the measured power range.

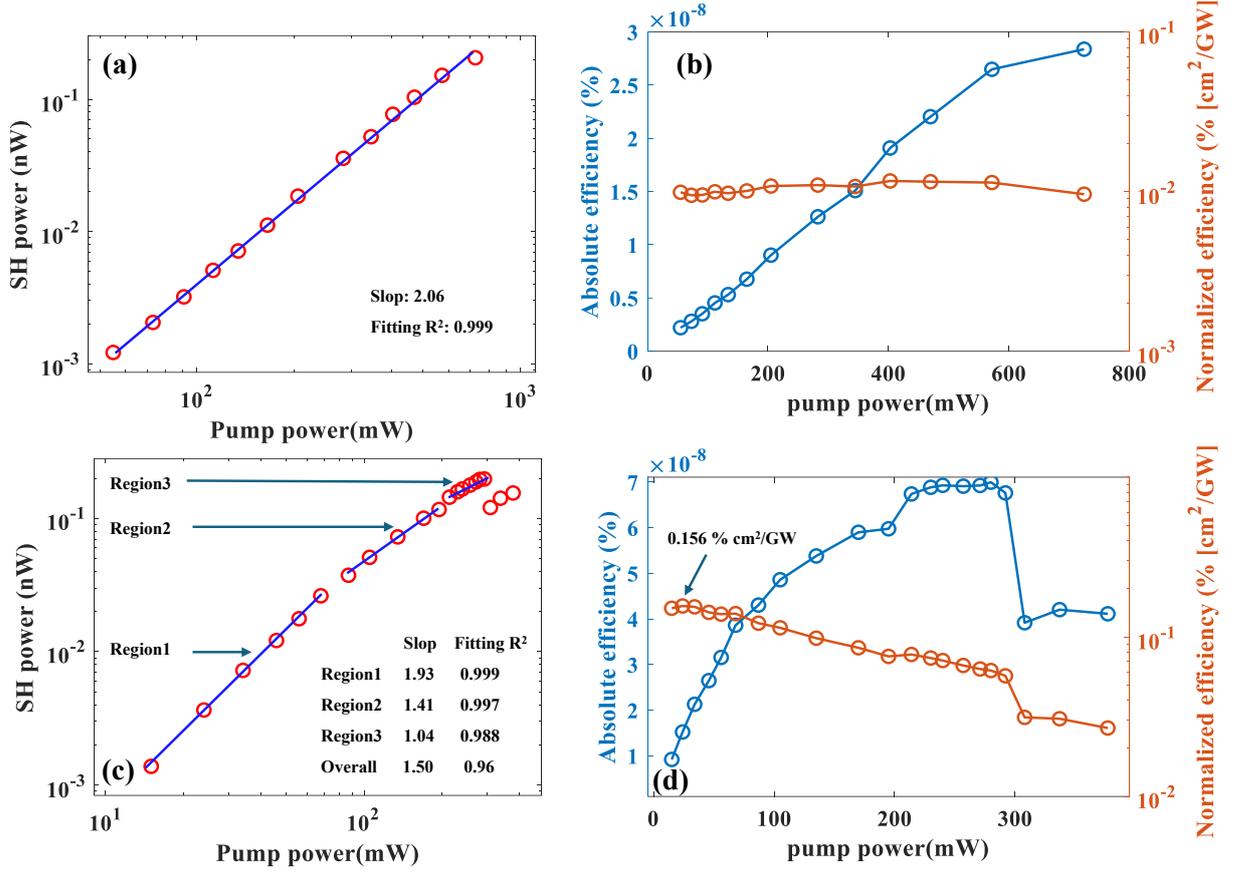

*Figure 4. SH power (a, c) and conversion efficiency (b, d), for the low-Q device (a, b) and high-Q device (c, d). The low-Q device shows a close to 2 logarithmic slope between the SH power and pump power up to high power, with almost constant normalized conversion efficiency. The high-Q device shows piecewise, and progressively decreased slope as pump power increases.*

In contrast, the high-Q device does not exhibit a single quadratic power dependence across the full pump range but instead shows piecewise scaling behavior. In the low-power regime (< 70 mW), the slope in the log-log plot remains close to 2, indicating conventional SHG-like scaling. In this regime, the normalized conversion efficiency is approximately one order of magnitude higher than that of the low-Q device, reaching 0.156 % $cm^2$/GW. As the pump power increases, however, the fitted slope decreases progressively, while the pump wavelength corresponds to the maximum SH signal shifts to longer wavelength, consistent with the power-dependent resonance red shift described above. Furthermore, at a pump power of around 300 mW, the SH power drops despite further increase in pump power. This behavior is accompanied by a blue shift in the pump wavelength corresponding to the maximum SH signal (Figure S2), leading to a reduction in the absolute conversion efficiency. These observations show that the SHG enhancement in the high-Q device is increasingly constrained at elevated pump powers by additional power-dependent effects. Their much weaker manifestation in the low-Q device suggests a strong connection to the resonant nature of the structure.

As noted above, the SHG data presented so far were obtained using a resonance-tracking protocol, in which, for each pump power, the pump wavelength was iteratively tuned around the resonance until the maximum SH signal was reached and then allowed to approach a quasi-steady level. We next investigate the transient SH response of the high-Q device to gain further insight into its nonideal power-dependent behavior. As shown in Figure 5, the high-Q device is initially kept in the dark ($t < 0$), and the SH signal is recorded after the pump is abruptly incident on the device ($t > 0$) at a preset power and wavelength, unless otherwise specified. For each measurement, the pump power was stabilized before illumination, and the pump was switched onto the device by lowering a flip mirror placed in front of the sample. The low-power resonance wavelength, determined from the FW transmission spectrum at a few milliwatts, is 1549.4 nm.

Figure 5a shows the transient SH response at a fixed pump power of 270 mW. Here, $\lambda_{\mathrm{maxSH}}$ = 1550.4 nm denotes the pump wavelength that yields the maximum SH output under the resonance-tracking protocol at this power. The blue horizontal trace represents the corresponding optimized quasi-steady SH level, which also contributes to one data point in Figure 4. When the device is illuminated directly from the dark at $\lambda_{\mathrm{maxSH}}$, the SH signal exhibits only a slow and relatively weak increase (green curve), remaining far below the optimized value. For red-detuned excitation, i.e., at $\lambda_{\mathrm{maxSH}} + 0.1$ nm, the response is even weaker. In contrast, for blue-detuned excitation at $\lambda_{\mathrm{maxSH}} - 0.3$ nm and $\lambda_{\mathrm{maxSH}} - 0.4$ nm, the SH signal first builds up with time and can temporarily exceed the optimized quasi-steady level before relaxing. For the more strongly blue-detuned case, $\lambda_{\mathrm{maxSH}} - 0.4$ nm, which is also closer to the low-power resonance, the rise occurs earlier and more rapidly, followed by a pronounced overshoot and subsequent decay. The response at $\lambda_{\mathrm{maxSH}} - 0.3$ nm is slower and reaches its maximum at a later time. However, when the pump is tuned further to the blue side, beyond the low-power resonance, to $\lambda_{\mathrm{max\ SH}} - 1.5$ nm, the transient behavior changes qualitatively: the SH signal shows an abrupt initial rise followed by a rapid decay toward a much lower level.

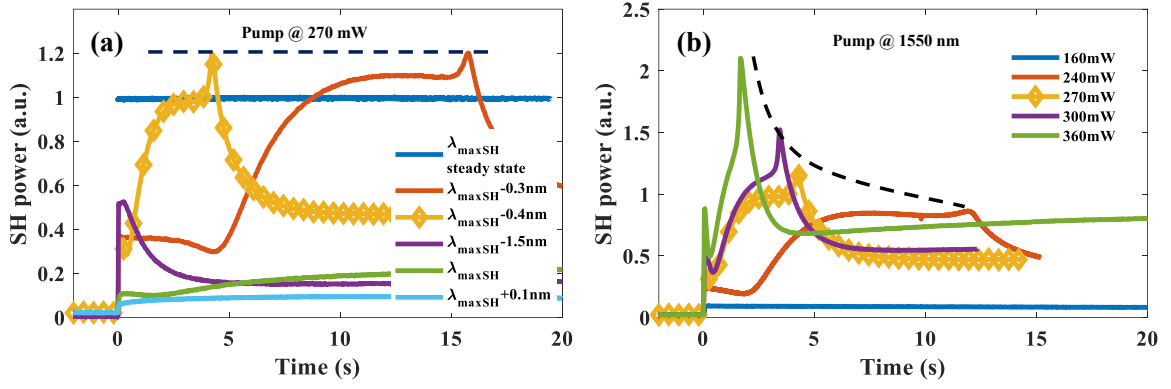

*Figure 5. Transient dynamic of the high-Q device SH response. (a) under constant pump power, and (b) under constant pump wavelength. $\lambda_{maxSH}$ denotes the pump wavelength that yields the maximum SH signal at quasi-steady state under 270 mW, as determined using the resonance-tracking protocol. The yellow curves with markers indicate the same measurement for direct comparison. The dashed lines are included only as guides to the eye.*

Figure 5b presents the transient SH response at a fixed pump wavelength of 1550 nm, while varying the pump power. At 160 mW, the SH signal shows an almost step-like onset and remains at a relatively low level. At 240 mW, a delayed increase appears before the signal reaches a broad plateau. As the pump power is further increased to 270, 300, and 360 mW, the transient response evolves into a more pronounced rise–overshoot–relaxation behavior: the onset becomes earlier, the growth becomes steeper, and the peak SH signal becomes larger. The yellow curve with markers in Figure 5b corresponds to the same measurement as the $\lambda_{\max SH} - 0.4$ nm trace in Figure 5a, included here for direct comparison.

The combined observations above suggest a phenomenological picture for the SHG process in the high-Q device. When the pump wavelength falls within the spectral range over which the resonance can still be excited (which may extend over several nanometers but only on the appropriate side according to the low-power resonance), part of the incident light couples into the device and generates a locally enhanced field. This field buildup can amplify even weak material absorption and is consistent with a thermo-optically induced red shift of the resonance[37–39]. As the instantaneous resonance moves toward the fixed pump wavelength from the shorter-wavelength side, the spectral overlap increases, enabling stronger coupling, larger field enhancement, and further resonance red shift. The closer the pump is to the dark-state resonance, and the higher the pump power, the faster this positive-feedback process develops. When the evolving resonance passes through the pump wavelength, a transient maximum in SH output is produced. Continued red shift then drives the system to the detuned side of the resonance, leading to a decay in SH power, before a relatively flat regime is reached where resonance shift and detuning approximately balance each other. A qualitatively similar transient response is also observed in the low-Q device (Figure S3), although the effect is less pronounced than in the high-Q device, consistent with its broader linewidth and weaker field enhancement.

The transient SH peak depends mainly on pump power (indicated by dashed lines), while a faster rise is accompanied by a faster decay, consistent with a delayed thermo-optic response. Further support for this interpretation comes from the resonance wavelength extracted under the resonance-tracking protocol (Figure S2b), which begins to exhibit a blue shift above a certain pump power, whereas no such feature is observed in the linear transmission sweep (Figure 3a). This difference indicates that the thermo-optic response is not instantaneous but requires time to build up. This phenomenological picture also explains the effectiveness of the resonance-tracking procedure used above: by slowly red tuning the pump in small steps, the measurement is more likely to keep the pump near the instantaneous resonance and thus recover the maximum SH signal at a given pump power.

## 3. Discussion and conclusion

Beyond the thermally induced resonance shift discussed above, the SHG results further suggest the presence of additional loss or degradation channels that progressively reduce the attainable SH output, preventing the normalized efficiency of the high-Q device from being maintained at its low-power level. The observations indicate that these channels are positively correlated with both the local field intensity in the resonant mode, set by the resonant field enhancement (Q) and the pump power; and the accumulated time under resonant excitation, pointing to a thermally assisted process. First, the low-Q device retains an approximately quadratic power dependence over a much wider pump range, whereas the high-Q device shows a slope close to 2 only at low power, suggesting that the degradation becomes more pronounced as the absolute local resonant field intensity increases. Second, in the intermediate-power regime, the transient SH peak recovers a slope close to 2 (Figure S4a) while the steady-state response is already strongly sub-quadratic (Figure 4c, region 2 and 3). Since the resonance-tracking protocol is designed to keep the device near the instantaneous optimum resonance condition, this difference indicates that an additional loss process builds up over time while the device remains under resonant excitation. Third, at higher pump powers (>400 mW), the transient optimum again starts to deviate from quadratic scaling, implying either that the resonance drift becomes too rapid for the true optimum to be captured or that the device undergoes further degradation at higher power. Finally, even in the intermediate-power regime where the slope extracted from transient response is close to 2, its normalized efficiency (Figure S4b) remains below that of the low-power regime (Figure 4d), indicating that, besides time-accumulated loss, the large local field itself is also associated with additional loss.

The exact origin of these processes remains an open question, but possible contributors include local-field-related mechanisms such as defect-state absorption in the SRN or LN[40,41], multiphoton absorption[42,43], or photorefraction in LN[44]; as well as thermally assisted processes such as heating-enhanced absorption and resonance degradation caused by nonuniform thermal accumulation[45,46]. Further investigation is needed both to clarify the device physics and to improve the CW-driven SHG performance of the present platform. Experimentally, chopped or modulated CW excitation may help probe and alleviate thermally assisted effects, while temperature-dependent measurements could isolate the thermal contribution more directly. From the design perspective, device geometries that produce different modal distributions and field localizations (e.g. using thicker LN to hold more optical mode) should help identify the material origin of the nonideal behavior and guide further optimization of the resonant enhancement.

More broadly, our results highlight a key issue for CW nonlinear metasurfaces: the conversion efficiency is not a purely static material or device parameter, but an operating-state-dependent quantity governed by resonance dynamics under illumination. As a result, the resonance condition inferred from a high-power linear transmission sweep does not necessarily represent the true optimum for nonlinear conversion, and the reported efficiency can depend on how that operating point is reached. This issue is expected to be especially pronounced in high-Q devices, where the same narrow linewidth that strengthens the nonlinear interaction also amplifies the sensitivity to power-dependent drift and degradation. Similar piecewise power scaling can in fact be recognized in previous reports under both CW[14] and pulsed excitation[23], although it has not been explicitly analyzed. These findings suggest that benchmarking of CW-driven resonant nonlinear metasurfaces should move beyond a single peak-efficiency metric to include power dependence, temporal evolution, and path-dependent operation.

In summary, we demonstrate continuous-wave second-harmonic generation in a transmission-mode, etchless lithium niobate metasurface with a silicon-rich nitride grating layer patterned by standard CMOS-compatible processes. The hybrid LN–SRN design enables strong GMR-induced field confinement (Q~2300), leading to clearly measurable SHG at sub-kW/cm$^2$ CW pumping. A normalized conversion efficiency of 0.156 % cm$^2$/GW is obtained in the low-power regime, which is highly competitive among reported LN-based SHG metasurfaces[47].

Beyond the efficiency itself, the device reveals a distinct dynamic operating regime under CW excitation, including power-dependent resonance drift, transient overshoot, nonideal scaling, and protocol-dependent SH response, highlighting that the performance of resonant nonlinear metasurfaces cannot be fully described by a static efficiency metric alone. In addition to previously demonstrated SH-wavelength tuning in metasurfaces through variation of the incident angle[25], our results further suggest that the CW SH output wavelength can also be tuned in a seemingly gapless manner through pump-power-dependent resonance dynamics. These results establish the present etchless LN platform as a promising route toward compact CW nonlinear photonic devices, while also emphasizing the importance of resonance dynamics in the operation and benchmarking of CW-driven nonlinear metasurfaces.

## 4. Methods

*Numerical Simulation*: The eigen mode simulation is conducted using Ansys Lumerical MODE. The periodic boundary condition is assigned in x-direction. The optical transmission spectrum and optical field enhancement are simulated by Ansys Lumerical RCWA.

*Sample Fabrications*: The device was fabricated from x-cut thin-film LN (310 nm) on SiO2 (~2 μm) platform, with bulk LN as substrate (~500 μm), (samples purchased from NANOLN). The SRN was deposited via plasma-enhanced chemical vapor deposition (PECVD) process, as detailed in our previous work[48]. $SiH_4$ and $NH_3$ were used as reaction gases. By controlling the ratio between the two, SRN with different loss and refractive index can be obtained (measured optical constant in Figure S1). An approximately 250 nm thick negative resist (maN-2403) was then spin-coated and patterned by e-beam lithography (Elionix). The film was etched using the reactive ion etching process (RIE) and finally underwent cleaning with oxygen plasma and piranha.

*Optical Characterization*: As described in Figure 2c, measurements were performed using a tunable continuous wave laser (Keysight 8164B) in the telecom regime. Fiber collimator (FC220fc-1550) was used to collimate light from a Single mode fiber, which was then focused by a f=200 mm lens (AC254-200C), leading to a spot size of ~177 μm diameter at its focal point. For nonlinear measurements, the tunable laser was amplified by an EDFA (IPG-EAD-2K-C). The transmitted light for both linear and nonlinear measurements were focused by an uncoated lens (LA1131) and directly fiber (M7L02, 400 μm) coupled to a spectrometer (Ocean Optics Flame-T-XR1-ES) and a femtowatt detector (New Focus 2151). The reflective and coupling loss were not corrected, hence the real SH power should be higher.

**Acknowledgements**

Z.C. acknowledges a scholarship from the center for nanoscience and nanotechnology of the Hebrew University. Samples were fabricated at the center for nanoscience and nanotechnology of the Hebrew University.

Funding: The work was supported by a grant from the Israeli-USA Binational Science Foundation (BSF, 2024059).

**Data Availability Statement**

The data that supports the findings of this study are available from the corresponding author upon reasonable request

**Supporting Information**

Supporting Information is Attached.

Supporting Information

# Continuous Wave Second Harmonic Generation from an Etchless Lithium Niobate Resonant Metasurface

*Zetian Chen\*, Noa Mazurski, and Uriel Levy\**

The Faculty of Science, The Center for Nanoscience and Nanotechnology, Institute of Applied Physics, The Hebrew University of Jerusalem, Jerusalem 91904, Israel

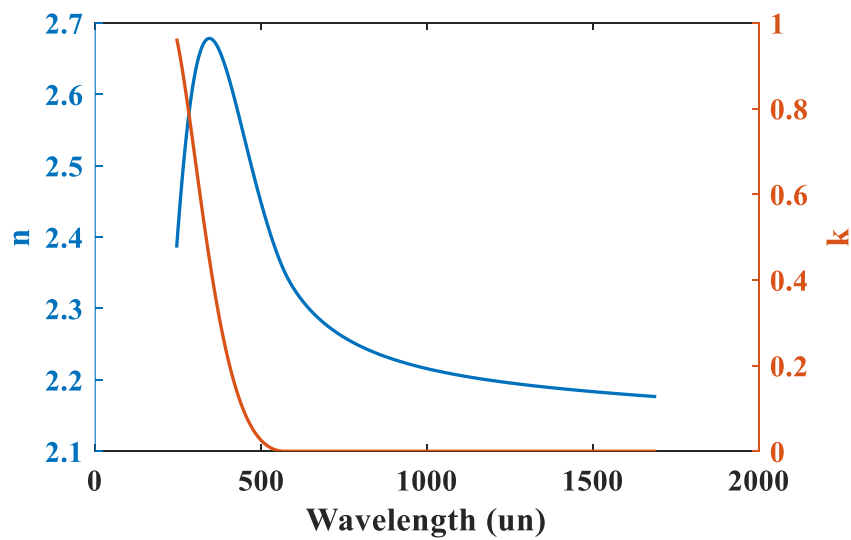

*Figure S1. The optical constants of the Silicon-rich nitride used in this work, measured by ellipsometer.*

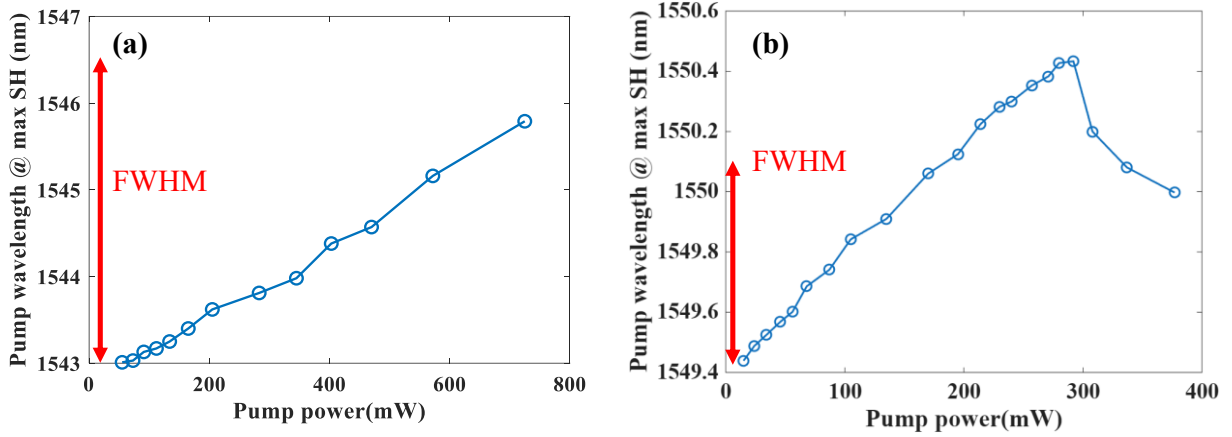

*Figure S2. Pump wavelengths at the maximum SH power under different pump powers for the (a) low-Q device, and (b) high-Q device. The red bars indicate the full-width-half-maximum (FWHM) of the resonance under low power sweep for each case. Compared to their FWHM, the high-Q device presents more pronounced shift than the low-Q one.*

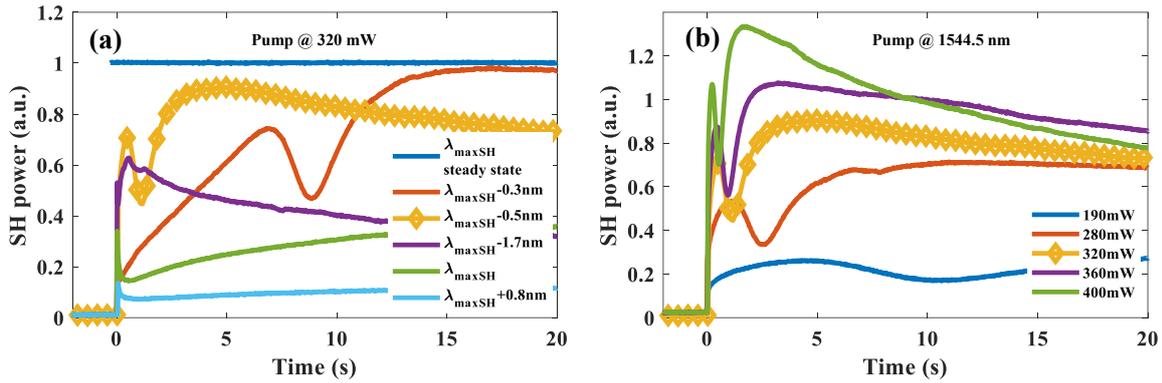

*Figure S3. The transient dynamic SH response of the low-Q device. (a) under constant pump power, and (b) under constant pump wavelength. $\lambda_{maxSH}$ in (a) corresponds to the pump wavelength that achieves maximum SH signal at quasi-steady state, followed by resonance-tracking protocol in the main text. The resonance wavelength extracted from low-power linear spectrum is at ~$\lambda_{maxSH}$-1.5 nm.*

The low-Q device exhibits qualitatively similar transient behavior, suggesting that the same dynamic resonance-shift mechanism is still presented. However, the transient overshoot is strongly suppressed, and the subsequent decay is slower than in the high-Q device. This behavior is consistent with the broader resonance linewidth of the low-Q device, which reduces the sensitivity of the SH output to pump-resonance detuning, together with its weaker resonant field enhancement, which is likely to reduce the magnitude and rate of the power-induced resonance shift.

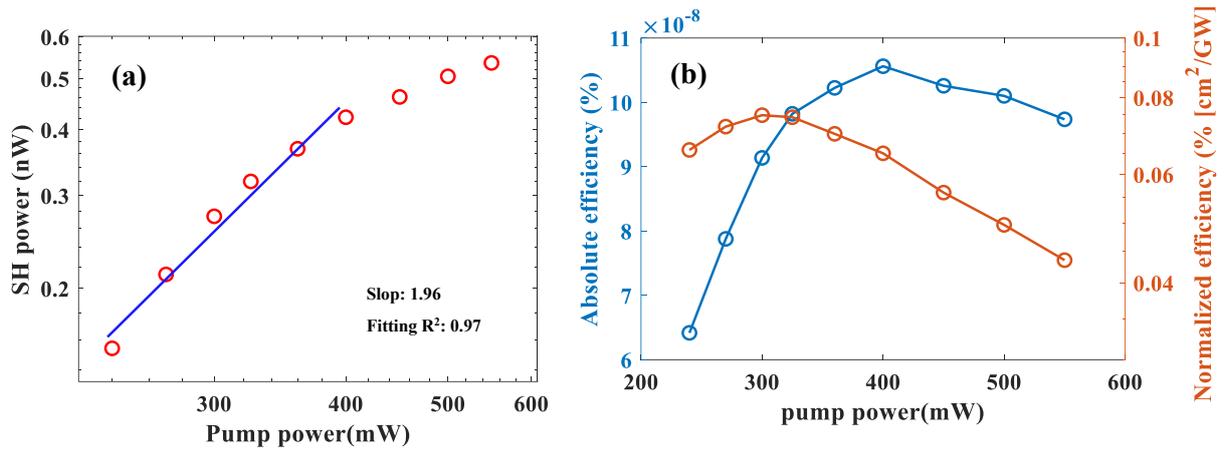

*Figure S4. (a) The transient SH power, extracted from the overshoot peak in Figure 5a in the main text, with respect to the pump power in log-log scale. The data below 240 mW is not recorded, since the rise-overshoot-decay dynamic is not activated. The slope is fitted up to 400 mW, above which a deterioration of the slope can be clearly seen. (b) The SH conversion efficiency extracted from (a).*